# Mental Task Classification Using Electroencephalogram Signal

Zeyu Bai
University of California, Los Angeles
405 Hilgard Avenue, 90095
Zeyubai21@engineering.ucla.

Ruizhi Yang
University of California, Los Angeles
405 Hilgard Avenue, 90095

Youzhi Liang
Massachusetts Institute of Technology
77 Massachusetts Avenue, 02139

*Abstract*— **This paper studies the classification problem on electroencephalogram (EEG) data of mental tasks, using standard architecture of three-layer CNN, stacked LSTM, stacked GRU. We further propose a novel classifier - a mixed LSTM model with a CNN decoder. A hyperparameter optimization on CNN shows validation accuracy of 72% and testing accuracy of 62%. The stacked LSTM and GRU models with FFT preprocessing and downsampling on data achieve 55% and 51% testing accuracy respectively. As for the mixed LSTM model with CNN decoder, validation accuracy of 75% and testing accuracy of 70% are obtained. We believe the mixed model is more robust and accurate than both CNN and LSTM individually, by using the CNN layer as a decoder for following LSTM layers. The code is completed in the framework of Pytorch and Keras. Results and code can be found at https://github.com/theyou21/BigProject.**

## I. INTRODUCTION

Deciphering mental activities shed light on the research on brain-computer interfaces and mental health with electroencephalogram (EEG) [1, 2]. Extensive studies applying machine learning techniques, such as networks and Support Vector Machines (SVM), have been transferred to the modelling using EEG data [3-7]. EEG, serving as a noninvasive measurement of brain activity., is broadly used to acquire the patterns of brain activity information [8-10]. In prior arts, emotion recognition, diagnosis of diseases, evaluation of medical treatment, brain-mobile phone interface, etc., have been investigated using EEG technology. [11-14].

### A. Data Description

Input data consists of 9 subjects of EEG signals, each with 288 trails sampled through 4 seconds by 250 Hz, resulting in a total of 1000 time signals. A total of 25 electrodes are placed on different locations on the top of head. Among the 25 channels, the last 3 are EOG channels, which are not used for data classification in this report. Combining all subjects and trails together gives a total of ((288*9), 22, 1000) input data. After removing all trails with any NAN values, we got the following training, validation and testing data sets:

- X train: (2358, 22, 1000), y train: (2358,)
- X val: (100, 22, 1000), y train: (100,)
- X test: (100, 22, 1000), y train: (100,)

By default, the validation and testing sets have 100 trails if not specifically mentioned. We can also perform implicit validation in the training process, which means validation set is randomly selected from the training set in each epoch. This approach can avoid the bias to a fixed validation set in the model selection.

### B. Model Configuration

Tackling this problem, the first thought would be using a convolutional neural network (CNN) model. CNN has been very successful in computer vision where the inputs usually have high dimensional features (each pixel of a image is a dimension). Therefore, it is suitable to deal with the long sequence data and is easy to be implemented. Because the dataset is small, we need to carefully choose the number of parameters so that overfitting does not happen on this small-scale data. On the other hand, it is also natural to try recurrent neural networks (RNN) which can handle temporal data through modeling the underlying dynamical behaviors 1 in the time sequence. Since the temporal data is consecutively generated from the same electrode, there is strong correlation in time within the sequence. Directly learning from the long sequence with noise in RNN models is usually challenging. Instead, using FFT to filter out the noise and down-sample the sequence should be helpful. This is an important pre-processing step in our RNN models. Finally, in order to take advantage of both CNN and LSTM, a mixed model where a CNN layer followed by LSTM layers is developed. It turns out the CNN layer acts as a very effective decoder for LSTM layers, and the mixed LSTM model has the highest validation and testing accuracy.

## II. RESULTS AND DISCUSSTION

### A. Three-Layer CNN Model

The size of input data is different from our familiar CIFAR-10 dataset, which has one more dimension compared to this given EEG data. Thus an 1-D convolution is done in time domain. This CNN model consists of a 1-D convolution layer, 1-D maxpooling layer and another 2 linear layers, as tabulated in 1. Batch-normalization are used before activation and dropout is used for regularization. Data mean subtraction is done in the dimension of examples.

A grid search of convolution filter number, filter size, pool size, batch size and learning rate is done for hyperparameter optimization. Optimization ranges are shown in table 1. There are more hyperparameter to study such as number of units in linear layer, stride for convolution layer, etc. However, due to the limit of our computation resource, only limited number of hyperparameter are investigated. Usually, it's not obvious to find a clear optimum space of hyperparameters. Thus we only present the results of hyperparameters with best validation accuracy and use this set of hyperparameter for testing.

Figure 1 shows the training loss, together with training and validation accuracy history. Validation accuracy achieves 72% and testing accuracy of the same model is 62%. Since hyperparameter optimization is done according to validation accuracy and k-fold cross validation is not used here due to the limit of CPU time, overfitting on the fixed validation set is possible. We believe the testing accuracy may improve by k-fold cross validation.

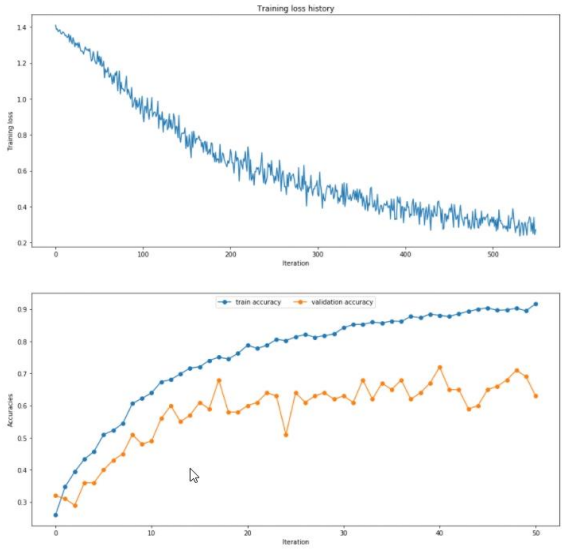

Figure 1. History of loss and accuracy of CNN model

### B. Stacked LSRM and Stacked GRU Models

The vanishing or exploding of loss gradient makes the vanilla RNN model difficult to train for long sequences. To deal with this problem, more sophisticated models such as long short-term memory (LSTM) and gated recurrent unit (GRU) models are invented [16,17]. In this report, we build the stacked LSTM and GRU models to deal with the EEG data. First, the time sequences are preprocessed and down-sampled through FFT. A high-pass filter above 5 Hz is applied to remove the long-period cycles which are likely to be introduced by eye motion or other disturbance [18]. The down-sampling is then performed through FFT, which can avoid aliasing error, remove high-frequency noise, and maximally maintain the feature of original sequence.

The LSTM and GRU models have the same architecture except the module used in the recurrent layers. Three recurrent layers are stacked, followed by a fully-connected layer and an output layer. Dropout is used in recurrent layers and the fully-connected layer. Batch normalization and ReLU activation are applied in the fully-connected layer. The cross-entropy loss with a softmax function is computed in the output layer. More details about the architecture can be found in table 2. In the training process, early-stopping is considered as a regularization and the model with the highest validation accuracy is recorded for testing.

Different down-sampling sizes are experimented. Figure 2 shows that small down-sampling sizes lead to better testing accuracies. Overall, the LSTM model performs better than GRU models. Due to the early-stopping strategy used in the training process, no severe overfitting is observed on small down-sampling sizes. Respectively, the LSTM model and GRU model achieve 55% and 51% testing accuracies with a down-sampling size of 50. Figure 3 shows a example of down-sampled data (red), and the output of the high-pass filter (blue) is plotted for comparison. It's clear that the down-sampled sequence can keep the trend fairly well although only 50 points are used. The training history of the LSTM model is shown in Figure 4. The model achieves best validation accuracy 51.22% after 74 epochs' training with training accuracy of 50.89%, indicating that the model isn't overfitting yet. Therefore, the observed 55% testing accuracy is reasonable although the test set is small.

Due to limited CPU time, the hyper parameters are optimized manually for each down-sampling size. Figure 2 shows that the overfitting is still the main problem for large down-sampling size. We believe that an extensive hyperparameter optimization can improve the performance given more computational resource.

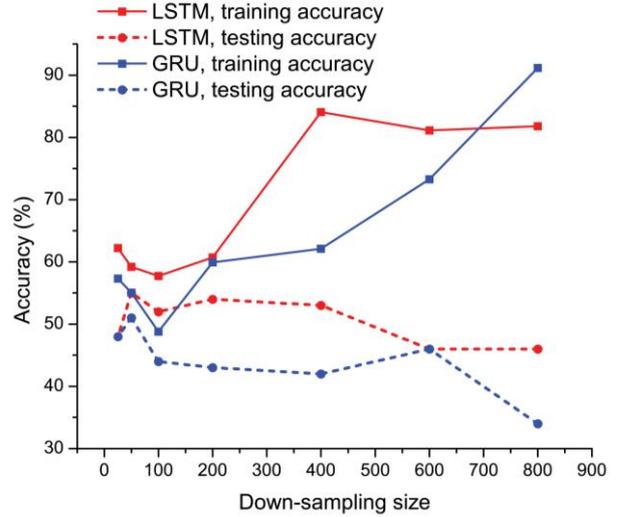

Figure 2. training and testing accuracy versus down-sampling size

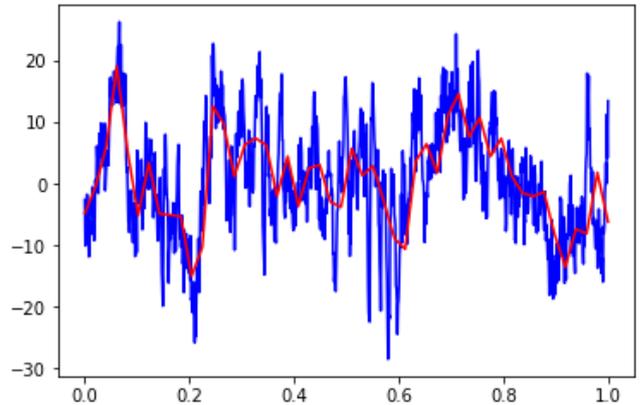

Figure 3. an example of down-sampling with 50 points

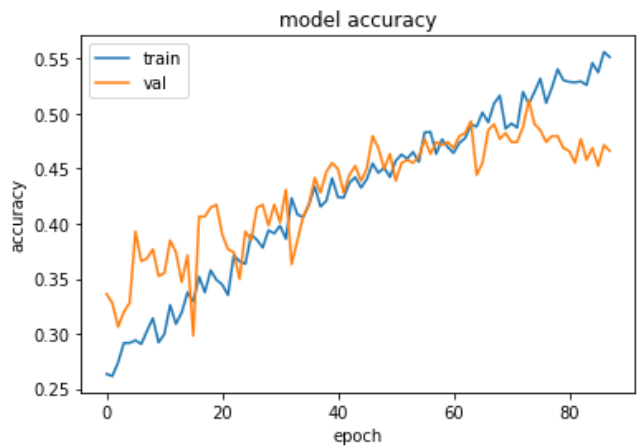

Figure 4. training accuracy and validation accuracy for the LSTM model with the down-sampling size of 50

## C. Mixed LSRM Model with CNN Decoder

In this section, we use a CNN layer as a decoder for original input data, followed by single or multiple LSTM layers to deal with the decoded temporal data. A fully connected layer is used for output, where the cross entropy loss with a softmax function is computed for classification.

Applying decoders to extract high-level features is very common in sequence processing. In fact, the filtering and down-sampling through FFT in the stacked LSTM or GRU models can also be considered as a decoder, which decodes the data into single shorter sequences by maintaining the mid-range frequency component. In the mixed LSTM model, we let CNN to extract multiple high-level features from the raw data by using a relatively large filter size and stride. Therefore, the sequence is decoded into multiple shorter sequence representing different high-level features.

With the 2-layer LSTM structure shown in table 3, this mixed LSTM model provides 70% validation accuracy and 71% testing accuracy. With 1-layer LSTM structure, the model achieves 75% validation accuracy and 70% testing accuracy. The 3-layer LSTM has similar accuracy on the basis that the total number of hidden units are the same with the previous two models. In all 3 cases, the testing accuracy is higher than CNN model or stacked LSTM and GRU models. We also tried LSTM with different hidden units to find the best hyperparameter. Details can be found in table 3.

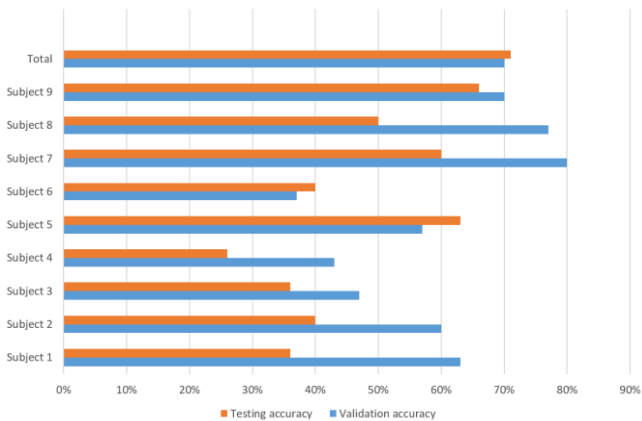

Figure 5. training and testing using different subjects of data

## D. Individual Subject Study

Using the 2-layer LSTM model, we also did a subject study across all 9 subjects. By training, validation and testing using their own data separately, Figure 5 shows validation and testing data of each subject, together with the total set of data. It is obvious that there exist big gaps between different subjects, and the total set of data has better performance than any individual ones. This is as expected since more data would certainly contributes to the model accuracy, while smaller data sets will cause high variance in the model performance.

## III. CONCLUSION

In this report, four different types of neural networks are developed - the 3-layer CNN, the stacked LSTM, the stacked GRU, and the mixed LSTM model with CNN decoder. All models achieve accuracies higher than 50%. Among them, the mixed model has best performance. The long sequence are fed into a CNN layer, which acts as a decoder and extract multiple high-level temporal features. These high level features are much easier for LSTM to deal with because of their shorter lengths. Therefore, the mixed LSTM model can combine the advantage of CNN and LSTM.

**Appendix 1: Method Three-Layer CNN Model**
We did hyperparameter optimization on CNN model as mentioned in the report, with the following range: • filter number: (30, 20, 10) • filter size: (28, 20, 12, 4) • pool size: (4, 2) • batch size: (200, 100, 50) • learning rate: ($1 \times 10^{-3}, 5 \times 10^{-4}, 1 \times 10^{-4}$)

| Layer | Note |
| --- | --- |
| Conv1d | filter_num=22, filter_size=28, stride=4 |
| Batchnorm | |
| Activation | Relu |
| Dropout | p=0.5 |
| MaxPool | pool_size=4, stride=4 |
| Flatten | |
| Linear | num_unit=20 |
| Activation | Relu |
| Linear | num_unit=4 |
| Activation | Softmax |

Table 1. CNN structure

Data preprocessing: (1) shuffling with the same random seed to forbid the influence of data recording order; (2) mean-subtraction on axis = 0 to normalize data from different trails. Optimizer: Adam Loss function: CrossEntropyLoss

**Stacked LSTM and Stacked GRU Models**
The architecture for stacked LSTM and Stacked GRU models are the same except the recurrent module (LSTM or GRU) used in the recurrent layers. The dropout and recurrent dropout probablility are changed according to the sequence length in the training to avoid overfitting. The stackted LSTM and GRU models also differ in dropout probablilities. The table shows the specific case of LSTM for down-sampling size of 50. Data preprocessing: (1) shuffling with the same random seed to forbid the influence of data recording order; (2) high-pass filter above 5 Hz to remove the long-period cycles; (3) down-sampling through FFT to avoid aliasing error (range of down-sampling size: 25 to 800). Optimizer: RMSProp Loss function: CrossEntropyLoss

| Layer | Note |
| --- | --- |
| LSTM/GRU | units_num=200, return_sequences=True |
| Dropout | p=0.6, recurrent_p=0.6 |
| Activation | tanh(output), sigmoid(recurrent) |
| LSTM/GRU | units_num=100, return_sequences=True |
| Dropout | p=0.5, recurrent_p=0.5 |
| Activation | tanh(output), sigmoid(recurrent) |
| LSTM/GRU | units_num=50, return_sequences=True |
| Dropout | p=0.4, recurrent_p=0.4 |
| Activation | tanh(output), sigmoid(recurrent) |
| Linear | num_unit=100 |
| Batchnorm | |
| Activation | Relu |
| Dropout | p=0.5 |
| Linear | num_unit=4 |
| Activation | Softmax |

Table 2. LSTM/GRU structure

Mixed LSTM Model with CNN Decoder The 1-D convolutional layer acts as a decoder with the output dimension of $246 = ((1000 − 20)/4 + 1)$. A total of 40 features are extracted by CNN layer, which are high-level temporal features of original sequence. Then, the LSTM layers can perform classification on these high-level features.

| Layer | Note |
| --- | --- |
| Conv1d | filter_num=40, filter_size=20, stride=4 |
| Batchnorm | |
| Activation | Relu |
| Dropout | p=0.5 |
| MaxPool | pool_size=4, stride=4 |
| LSTM | units_num=30, return_sequences=True |
| Batchnorm | |
| Activation | tanh(output), sigmoid(recurrent) |
| Dropout | p=0.5 |
| LSTM/GRU | units_num=20, return_sequences=True |
| Batchnorm | |
| Activation | tanh(output), sigmoid(recurrent) |
| Dropout | p=0.5, recurrent_p=0.5 |
| Flatten | |
| Linear | units_num=4 |
| Activation | Softmax |

Table 3. Combine structure

Data preprocessing: (1) shuffling with the same random seed to forbid the influence of data recording order; (2) mean-subtraction on axis=1(time sequence dimension). Optimizer: Adam Loss function: CrossEntropyLoss

**Appendix 2: Performance for Algorithms**
Notation:
CNN (a) (b): CNN with filter num=a and filter size=b
LSTM t(a): LSTM with down-sampling size of a
GRU t(a): GRU with down-sampling size of a
MixLSTM (a): CNN layer followed by a LSTM layers

| Algorithm | Training accuracy(%) | Validation accuracy(%) | Testing accuracy(%) |
|---|---|---|---|
| CNN_30_28 | 91.68 | 72.0 | 62.0 |
| CNN_30_20 | 93.59 | 71.0 | 60.0 |
| CNN_30_12 | 92.23 | 69.0 | 59.0 |
| CNN_30_4 | 88.08 | 73.0 | 54.0 |
| CNN_20_28 | 91.56 | 68.0 | 62.0 |
| CNN_20_20 | 89.69 | 67.0 | 67.0 |
| CNN_20_12 | 89.35 | 71.0 | 60.0 |
| CNN_20_4 | 84.98 | 67.0 | 59.0 |
| CNN_10_28 | 81.85 | 68.0 | 58.0 |
| CNN_10_20 | 80.44 | 68.0 | 69.0 |
| CNN_10_12 | 77.52 | 70.0 | 57.0 |
| CNN_10_4 | 73.74 | 67.0 | 59.0 |

| Algorithm | Training accuracy(%) | Validation accuracy(%) | Testing accuracy(%) |
|---|---|---|---|
| LSTM_t25 | 62.21 | - | 48.0 |
| LSTM_t50 | 59.19 | - | 55.0 |
| LSTM_t100 | 57.73 | - | 52.0 |
| LSTM_t200 | 60.70 | - | 54.0 |
| LSTM_t400 | 84.05 | - | 53.0 |
| LSTM_t600 | 81.12 | - | 46.0 |
| LSTM_t800 | 81.81 | - | 46.0 |
| GRU_t25 | 57.32 | - | 48.0 |
| GRU_t50 | 55.04 | - | 51.0 |
| GRU_t100 | 48.78 | - | 44.0 |
| GRU_t200 | 59.93 | - | 43.0 |
| GRU_t400 | 62.12 | - | 42.0 |
| GRU_t600 | 73.27 | - | 46.0 |
| GRU_t800 | 91.17 | - | 34.0 |

| Algorithm | Training accuracy(%) | Validation accuracy(%) | Testing accuracy(%) |
|---|---|---|---|
| MixLSTM_1 | 96.4 | 75.0 | 70.0 |
| MixLSTM_2 | 87.8 | 70.0 | 71.0 |
| MixLSTM_2, units_num in LSTM = (10,10) | 71.67 | 62.0 | 67.0 |
| MixLSTM_3 | 80.79 | 62.0 | 70.0 |